\documentstyle[preprint,aps,epsf]{revtex}
\begin{document}
\draft \preprint{MKPH-T-97-8} 
\title{ Unitarity constraint for threshold coherent pion
  photoproduction on the deuteron and chiral perturbation theory}
\author{P.\ Wilhelm} 
\address {Institut f\"ur Kernphysik, Johannes Gutenberg-Universit\"at
  Mainz, D-55099 Mainz, Germany\\ Electronic address:
  wilhelmp@kph.uni-mainz.de}
\date{\today}
\maketitle
\begin{abstract}
  The contribution of the two-step process $\gamma d\rightarrow
  pn\rightarrow\pi^0 d$ to the imaginary part of the amplitude for
  coherent pion production on the deuteron is calculated exploiting
  unitarity constraints.  The result shows that this absorptive
  process is not negligible and has to be considered in an extraction
  of the elementary neutron production amplitude from the $\gamma
  d\rightarrow\pi^0 d$ cross section at threshold.  In addition, it is
  argued that a consistent calculation of $\gamma d\rightarrow\pi^0 d$
  in baryon chiral perturbation theory beyond next-to-leading order
  requires the inclusion of this absorptive process.
\end{abstract}
\narrowtext 
\newpage

Recently there has been considerable interest in the coherent
electromagnetic production of pions from the deuteron near threshold.
The main motivation thereby is to gain information on the elementary
neutron amplitude $\gamma n\rightarrow\pi^0 n$ which is experimentally
not directly accessible.  A first measurement of the $ed\rightarrow e
\pi^0 d$ reaction near $q^2_\mu=-0.075\,\mbox{GeV}^2/c^2$ will be
performed soon at MAMI \cite{mainz}. However, it is already known for
a long time that at threshold the $\gamma d\rightarrow\pi^0 d$ process
is dominated by two-nucleon production mechanisms \cite{koch}.
Therefore, a careful theoretical analysis which allows the separation
of the one-nucleon process is essential.  Recently, chiral
perturbation theory ($\chi$PT) in the heavy baryon formulation has
been applied to this reaction by Beane et al.\ \cite{beane} predicting
a real threshold amplitude.  Very recently, this work has been
improved and extended beyond next-to-leading order in the chiral power
counting scheme \cite{beanenew}.

It is the purpose of the present paper to point out that an additional
contribution of an absorptive two-step process, $\gamma d\rightarrow
pn\rightarrow\pi^0 d$, has to be included, which leads to a complex
amplitude even at threshold.  The presence of such a competing
deuteron disintegration channel has an analogue in the case of $\pi d$
elastic scattering, where the contribution of the absorptive process
$\pi d\rightarrow NN\rightarrow\pi d$ is known to be of the order of
10\% of the total amplitude (see e.g.\ \cite{ericson}). An effect of
this size would not be negligible for the electromagnetic reaction due
to the relative smallness of the single-nucleon amplitude, one is
mainly interested in.  In the case of $\pi d\rightarrow\pi d$, the
imaginary part of the scattering length $a_{\pi d}$ is related to the
total absorption cross section through
\begin{equation}
  \label{apid} \Im m\, a_{\pi d} = \frac{1}{4\pi} \lim_{p_{\pi
  d}\rightarrow 0}\, p_{\pi d}\,\,\sigma(\pi d\rightarrow X),
\end{equation}
where $p_{\pi d}$ is the pion momentum in the c.m.\ system.  This
relation follows directly from the optical theorem.  An analogous
unitarity constraint on the $\gamma d\rightarrow\pi^0 d$ amplitude
near threshold is present and will be treated as first point below.
In view of the fact that the absorptive process has not been
considered in \cite{beane,beanenew}, we will analyze its role within
the $\chi$PT framework as second point.

As is well known, the unitarity of the $S$-matrix leads to constraints
for the corresponding reaction amplitudes.  For our purposes, it is
sufficient to consider coupled two-particle channels where the
particles are subject to interactions which are invariant under
time-reversal. In this case, following the conventions of Ref.\
\cite{pilkuhn}, the imaginary part of the partial wave $T$-matrix
element of total angular momentum $J$ for the reaction
$a=(a_1a_2)\rightarrow b=(b_1b_2)$ fulfils the following unitarity
constraint
\begin{equation}
  \label{eq:uni}
  \Im m\, 
  T^J_{\lambda_{a_1}\lambda_{a_2}\rightarrow\lambda_{b_1}\lambda_{b_2}}(W) =
  \sum_c p_c \sum_{\lambda_{c_1}\lambda_{c_2}}
  T^J_{\lambda_{a_1}\lambda_{a_2}\rightarrow\lambda_{c_1}\lambda_{c_2}}(W)
  \left(
  T^J_{\lambda_{b_1}\lambda_{b_2}\rightarrow\lambda_{c_1}\lambda_{c_2}}(W)
  \right)^\ast.
\end{equation}
Here, $\lambda_{c_i}$ is the helicity of particle $i$ in the channel
$c$, and $p_c$ denotes the c.m.\ momentum of the two particles in the
channel $c$.  The first sum on the rhs of (\ref{eq:uni}) runs over all
open channels $c$ for a given total c.m.\ energy $W$.  In terms of the
$T^J$ the total helicity amplitude for $a\rightarrow b$ is given by
\begin{equation}
  \label{helamp}
  T_{\lambda_{a_1}\lambda_{a_2}\rightarrow\lambda_{b_1}\lambda_{b_2}} 
  (W,\phi,\theta) =
  8\pi W \sum_J (2J+1) e^{i(\lambda_a-\lambda_b)\phi}
  d^J_{\lambda_a\lambda_b}(\theta)
  T^J_{\lambda_{a_1}\lambda_{a_2}\rightarrow\lambda_{b_1}\lambda_{b_2}}(W),
\end{equation}
with $\lambda_a=\lambda_{a_1}-\lambda_{a_2}$,
$\lambda_b=\lambda_{b_1}-\lambda_{b_2}$, and $\theta$, $\phi$ are the
spherical coordinates of the outgoing particle $b_1$.

We apply relation (\ref{eq:uni}) to the three coupled channels $\pi^0
d$, $pn$, and $\gamma d$ at the $\pi^0$ production threshold, i.e., in
the limit $W\rightarrow W_0\equiv m_{\pi^0}+m_d$, which, of course,
implies $p_{\pi^0d}\rightarrow 0$. In this limit, therefore, no
contribution from $c=(\pi^0 d)$ to the sum occurs due to the vanishing
phase space factor.  Moreover, at threshold we can restrict ourselves
to partial waves of angular momentum and parity $J^\pi=1^-$.  Thus
$\pi^0 d\rightarrow pn$ is described by a single matrix element $A$,
with
\begin{equation}
  T^1_{\lambda_d\rightarrow\lambda_p\lambda_n} = \frac{1}{3}
  (101\lambda_p-\lambda_n|1\lambda_p-\lambda_n) A.
\end{equation}
For $\gamma d\rightarrow\pi^0 d$ at threshold, two independent matrix
elements remain, $E_{\pi^0d}$ and $M_{\pi^0d}$. They correspond to the
electric dipole (E1) and magnetic quadrupole (M2) radiation allowed
for the $1^+\rightarrow 1^-$ transition of the hadronic system. One
finds
\begin{equation}
  T^1_{\lambda_\gamma\lambda_d\rightarrow\lambda_d^\prime} =
  \frac{1}{3\sqrt{6}} \sum_{L=1,2} \sqrt{2L+1}\, (1-\lambda_d
  L\lambda_\gamma | 1\lambda_\gamma-\lambda_d)\,
  (\delta_{L,1}E_{\pi^0d}+\lambda_\gamma\delta_{L,2}M_{\pi^0d}).
\end{equation}
Finally, for $\gamma d\rightarrow {^3}P_1(pn)$ one obtains
\begin{eqnarray}
  T^1_{\lambda_\gamma\lambda_d\rightarrow\lambda_p\lambda_n} &=&
  \frac{1}{3\sqrt{2}} (101\lambda_p-\lambda_n|1\lambda_p-\lambda_n) 
  \nonumber\\
  & & \sum_{L=1,2} \sqrt{2L+1}\, (1-\lambda_d L\lambda_\gamma |
  1\lambda_\gamma-\lambda_d)\,
  (\delta_{L,1}E_{pn}+\lambda_\gamma\delta_{L,2}M_{pn}),
\end{eqnarray}
where $E_{pn}$ and $M_{pn}$ are the corresponding E1 and M2 matrix
elements of the disintegration process.  The various total cross
sections in terms of these matrix elements are
\begin{eqnarray}
  && \sigma(\pi^0 d\rightarrow pn) =
  \frac{4\pi}{3}\frac{p_{pn}}{p_{\pi^0d}} |A|^2,
  \label{eq:toth}\\
  && \sigma(\gamma d\rightarrow\pi^0 d) =
  \frac{4\pi}{6}\frac{p_{\pi^0d}}{p_{\gamma d}} \left( |E_{\pi^0d}|^2
    + |M_{\pi^0d}|^2 \right),
  \label{eq:totpi}\\
  && \sigma\left(\gamma d\rightarrow {^3}P_1(pn);\mbox{E1+M2}\right) =
  \frac{4\pi}{6}\frac{p_{pn}}{p_{\gamma d}} \left( |E_{pn}|^2 +
    |M_{pn}|^2 \right).
  \label{eq:totpn}
\end{eqnarray}

Taking now $a=(\gamma d)$ and $b=(\pi^0 d)$, relation (\ref{eq:uni})
leads to
\begin{equation}
  \label{central} \Im m\, E_{\pi^0d}(W_0) = \frac{1}{\sqrt{3}}\,
  p_{pn}\, E_{pn}(W_0)\, A^\ast(W_0),
\end{equation}
and an analogous relation for $M_{\pi^0d}$ and $M_{pn}$, respectively.
Since the lhs of (\ref{central}) is real, relation (\ref{central})
implies that the phases of the complex matrix elements $A$ and
$E_{pn}$ are equal. Indeed, evaluating (\ref{eq:uni}) with $a=(\pi^0
d)$, $b=(pn)$ and $a=(\gamma d)$, $b=(pn)$ provides us with
\begin{equation}
  \label{aphase} A(W_0) =
  |A(W_0)|\,\exp\left(i\delta_{{^3\!}P_1}(W_0)+in\pi\right),
\end{equation}
and
\begin{equation}
  \label{watson} E_{pn}(W) = |E_{pn}(W)|
  \,\exp\left(i\delta_{{^3\!}P_1}(W)+ik\pi\right), \quad W\leq W_0,
\end{equation}
respectively, where $\delta_{{^3\!}P_1}$ is the nucleon-nucleon
scattering phase shift in the ${^3}P_1$ channel.  Eq.\ (\ref{watson})
is simply Watson's theorem applied to deuteron photodisintegration and
is valid for all energies below the pion production threshold, whereas
Eq.\ (\ref{aphase}) is valid for $W=W_0$ only.  Finally, we mention
that taking $a=b=(\pi d)$ in (\ref{eq:uni}), leads to the constraint
for the imaginary part of the $\pi d$ scattering length in
(\ref{apid}).

Using (\ref{eq:toth}) and (\ref{eq:totpn}), and the detailed balance
relation
\begin{equation}
  \label{balance} 3 \,p_{\pi^0d}^2\, \sigma(\pi^0 d\rightarrow pn) = 4
  \,p_{pn}^2\, \sigma(pn\rightarrow\pi^0 d),
\end{equation}
our main result (\ref{central}) can be rewritten as 
\begin{equation}
  \left|\Im m\, E_{\pi^0d}\right| = \frac{1}{\sqrt{2}\,\pi} \,
  p_{pn}\, \sqrt{\frac{p_{\gamma d}}{p_{\pi^0d}}\, \sigma\left(\gamma
  d\rightarrow {^3}P_1(pn);\mbox{E1}\right)\,
  \sigma(pn\rightarrow\pi^0 d) }.
\end{equation}
At this level, there is, however, no way to fix the sign. In order to
get a numerical value, we take for the hadronic cross section the
experimental result given by Hutcheon et al.\ \cite{hutcheon},
\begin{equation}
  \lim_{p_{\pi^0d}\rightarrow 0} 2\,\frac{m_\pi}{p_{\pi^0d}}\, 
  \sigma(pn\rightarrow\pi^0 d) = 
  184\pm5\pm13\,\mu b.
\end{equation}
The partial cross section $\gamma d\rightarrow {^3}P_1(pn)$ is at
present not available although it could in principle be obtained from
a multipole analysis.  There are, however, reliable theoretical models
available which reproduce all deuteron photodisintegration data in
this energy region \cite{ha}.  At this energy, the E1 matrix element
is mainly given by $\pi$-exchange current contributions, which can be
calculated in a largely model-independent way by taking advantage of
gauge-invariance constraints (Siegert's theorem).  The underlying
nucleon-nucleon interaction and also all model-dependent transverse
electromagnetic currents, like the $\Delta$(1232) excitation current,
have little effect on this matrix element.  We take the value
\begin{equation}
  \label{e1} \sigma\left(\gamma d\rightarrow
  {^3}P_1(pn);\mbox{E1}\right) = 10.5\,\mu b
\end{equation}
from an updated version of the model of Ref.\ \cite{myletter}.
In order to relate our result to the $\chi$PT calculations of Beane et
al.\ \cite{beanenew}, we switch to their normalization (and notation)
of the electric dipole amplitude, $E_d\equiv E_{\pi^0 d}/4$, and obtain
\begin{equation}
  \label{result} \left| \Im m\,E_d\right| = 0.22\times
  10^{-3}/m_{\pi^+}.
\end{equation}
The cross section $\sigma\left(\gamma d\rightarrow
  {^3}P_1(pn);\mbox{M2}\right)$ is more than two orders of magnitudes
smaller than (\ref{e1}) and leads to $\left| \Im m\,M_d\right| =
0.018\times 10^{-3}/m_{\pi^+}$, where $M_d\equiv M_{\pi^0 d}/4$.
Nevertheless, the role of the M2 transition in the coherent production
at threshold deserves a more detailed investigation which will be
presented elsewhere. The main reasons are: ({\it i}) it provides an
additional possibility to test theoretical predictions (an
experimental separation of E1 and M2 requires a polarized deuteron
target), and ({\it ii}) the relative importance of the M2 transition
grows with the momentum transfer in the electroproduction process.

The result (\ref{result}) has to be compared with the value calculated
in \cite{beanenew},
\begin{equation}
  \label{chpt} E_d^{\chi PT} = 0.38 E_{0+}^{\pi^0n} - 2.6\times
  10^{-3}/m_{\pi^+} = -1.8\times 10^{-3}/m_{\pi^+},
\end{equation}
where the latter value is obtained taking the $\chi$PT prediction of
\cite{meissner} for the neutron electric dipole amplitude,
$E_{0+}^{\pi^0n}=2.13\times 10^{-3}/m_{\pi^+}$.  Thus the threshold
cross section itself is affected by less than 2\% only by this
imaginary part (\ref{result}).  However, there is no reason at all to
assume that the contribution of the absorptive process to the real
part of the amplitude is much smaller in magnitude than the imaginary
part.  Unfortunately, unitarity does not allow to estimate it.  It can
only be calculated within a model which to our knowledge has not yet
been done.  For the moment, in order to get a rough idea, one may look
into the in many respects analogous situation for $\pi d$ elastic
scattering.  There, three-body calculations suggest for the absorptive
contribution to the scattering length, $\Re e\,a_{\pi d}^{abs}\approx
-\Im m\,a_{\pi d}$ (see \cite{ericson} for an overview).  Assuming,
therefore $\left|\Re e\,E_d^{abs}\right|=0.22\times
10^{-3}/m_{\pi^+}$, one has to conclude that the neglect of the
absorption process in an analysis of the $\gamma d\rightarrow\pi^0 d$
cross section based on (\ref{chpt}) would lead to a systematic error
of the order of $\delta E_{0+}^{\pi^0n}=\pm 0.6\times
10^{-3}/m_{\pi^+}$ for the neutron amplitude.  This rough argument at
least demonstrates that a calculation of the absorptive contribution
is necessary before definite conclusions on the neutron amplitude can
be drawn.  One way to do this is to combine conventional models for
deuteron photodisintegration \cite{ha} with those for the pionic
disintegration of the deuteron \cite{niskanen}.

\begin{figure}
\centerline{\epsfxsize=10cm\epsffile{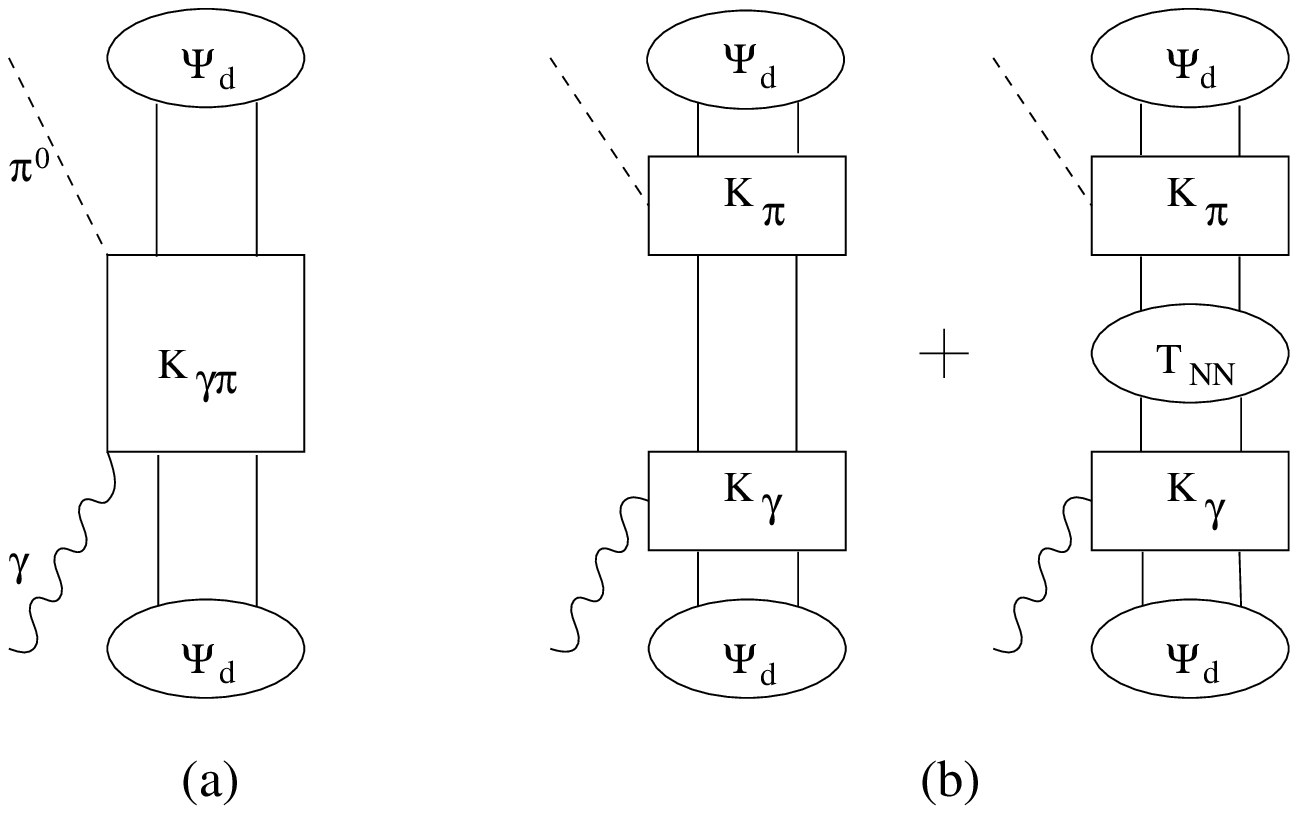}}
\caption{
  Schematical representation of the $\gamma d\rightarrow \pi^0
  d$ amplitude at threshold in the $\chi$PT framework: the
  non-absorptive part (a) is based on the irreducible $\gamma pn
  \rightarrow \pi^0 pn$ kernel $K_{\gamma\pi}$, and the absorptive
  part (b) combines the irreducible kernels $K_{\gamma}$ and $K_{\pi}$
  for $\gamma pn \rightarrow pn$ and $pn \rightarrow \pi^0 pn$,
  respectively.  $\psi_d$ is the deuteron wave function and $T_{NN}$
  the nucleon-nucleon scattering matrix in the ${^3}P_1$ partial wave.
}
\label{fig1}
\end{figure}

As second point we would like to address the question what is the role
of this absorptive process in the framework of heavy baryon $\chi$PT.
The $\chi$PT treatment of the process $\gamma d\rightarrow\pi^0 d$ is
sketched in Fig.\ \ref{fig1}.  The non-absorptive contribution, Fig.\
\ref{fig1}(a), is based on the two-nucleon irreducible kernel
$K_{\gamma\pi}$ for the $\gamma pn \rightarrow \pi^0 pn$ subprocess.
It is obtained from the effective chiral Lagrangian, and then
sandwiched between deuteron wave functions $\psi_d$.  $K_{\gamma\pi}$
sums all time-ordered diagrams which do not contain pure two-nucleon
intermediate states.  It has been calculated in \cite{beanenew} up to
and including all terms of order $\nu\leq 0$ of the expansion in terms
of powers of small momenta $(Q/\Lambda)^\nu$ where typically $Q\sim
m_\pi$, $\Lambda\sim m_N$.

The proper treatment of the absorptive contribution is shown in Fig.\
\ref{fig1}(b).  Formally it is given by
\begin{equation}
  \label{abs} T^{\chi PT/abs} = \psi_d^\dagger\, K_{\pi} G_0
  K_{\gamma} \,\psi_d + \psi_d^\dagger\, K_{\pi}
  G_0T_{NN}(W_0+i\epsilon)G_0 K_{\gamma} \,\psi_d.
\end{equation}
Here, the input from $\chi$PT are the two-nucleon irreducible kernels
$K_{\gamma}$ and $K_{\pi}$ for $\gamma pn \rightarrow pn$ and $pn
\rightarrow \pi^0 pn$, respectively.  These are linked by either the
free two-nucleon propagator,
\begin{equation}
  \label{prop}
  G_0 = \left( W_0-2m_N-\vec p^{\,2}/m_N+i\epsilon \right)^{-1},
\end{equation}
or the propagation via $G_0T_{NN}G_0$ which includes the
nucleon-nucleon interaction through the full off-shell $T$-matrix in
the ${^3}P_1$ partial wave. $T_{NN}$ (and $\psi_d$) has to be
calculated by solving the Lippmann-Schwinger equation for a
two-nucleon potential. Ideally, the potential is thereby also taken
from $\chi$PT as the sum of all irreducible $NN\rightarrow NN$
diagrams.

\begin{figure}
\centerline{\epsfxsize=8cm\epsffile{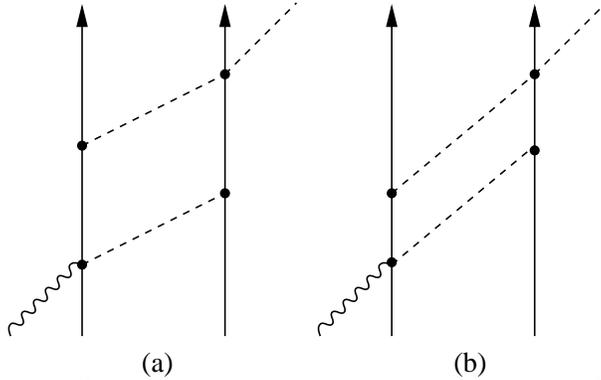}}
\caption{
  Two diagrams to illustrate our power counting arguments.  The
  time-ordered diagram (a) is part of the absorptive process, whereas
  diagram (b) although built up from the same vertices contributes to
  the non-absorptive part due to a different time-ordering. Solid,
  dashed, and wavy lines represent nucleons, pions, and photons,
  respectively.
}
\label{fig2}
\end{figure}

Now the question arises, where the absorptive contribution fits into
the $\chi$PT power counting scheme. We will answer it by means of
Fig.\ \ref{fig2}.  The time-ordered graph of Fig.\ \ref{fig2}(a) shows
an absorptive contribution which is contained in the first term of
(\ref{abs}).  The diagram of Fig.\ \ref{fig2}(b) is a part of
$K_{\gamma\pi}$ build up from the same interaction vertices as (a).
According to Weinberg's power counting rules \cite{wein1,wein2,wein3},
irreducible graphs with $N$ nucleons, $C$ separate connected pieces,
$L$ loops, and $V_i$ vertices of type $i$ contribute to order $Q^\nu$
with $\nu$ given by
\begin{equation}
  \label{nu}
  \nu=4-N-2C+2L+\sum_i V_i\Delta_i,
\end{equation}
where the index $\Delta_i$ is bounded by chiral invariance.  For pure
hadronic vertices one has $\Delta_i\geq 0$, while for vertices with
one photon $\Delta_i\geq -1$.  For the graph of Fig.\ \ref{fig2}(b)
with $N=2$, $C=L=1$ this leads to $\nu_{2(b)}\geq 1$.  Consequently
these type of contributions could be ignored by Beane et al.\ 
\cite{beanenew} in order to be consistent up to and including $\nu\leq
0$.

However, as has been stressed by Weinberg, graphs with intermediate
states containing {\em only} nucleons violate the simple power
counting rules, because of the small (nearly infrared-divergent)
energy denominators associated with the propagation of these states,
see Eq.\ (\ref{prop}).  Fig.\ \ref{fig2}(a) belongs to this class of
reducible graphs. In time-ordered perturbation theory, energy
denominators of states with at least one pion are of order $Q$, while
those with nucleons only are of order $Q^2/m_N$, when one assumes that
all momenta are of order $Q$.  This leads to count the order of Fig.\ 
\ref{fig2}(a) as $\nu_{2(a)}=\nu_{2(b)}-1 \geq 0$.

It could be objected, that the momenta of the intermediate particles
in Fig.\ \ref{fig2}(a) are not of order $Q$ but rather of order
$P\sim\sqrt{m_\pi m_N}$ which implies that the two-nucleon energy
denominator is of order $Q^{-1}$ rather than $Q^{-2}$.  Actually this
is true for the imaginary part of the diagram which just arises from
the kinematical situation where the nucleon momenta are equal to the
on-shell momentum $p_{pn}$ which is given by $p_{pn}^2/m_N=m_\pi$
applying nonrelativistic kinematics.  In such a situation also the two
intermediate pion momenta must be of order $P$ in order to change the
relative nucleon momenta of order $Q$, provided by the deuteron wave
function.  However, as will be seen soon, counting all powers of $P$
in the graph (a) of Fig.\ \ref{fig2} will not change the above
conclusion that it contributes already to the order $Q^0$. The
necessity to consider a modified power counting (due to the kinematics
of the reaction) was first noted by Cohen et al.\ \cite{friar},
studying the reaction $pp\rightarrow \pi^0pp$ near threshold.  In
order to get the worst case, we assume from now on that all vertices
in Fig.\ \ref{fig2}(a) arise always from the leading terms of the
chiral Lagrangian.  Following the steps in \cite{wein1} one counts for
Fig.\ \ref{fig2}(a): $P^3$ from three derivative couplings, $P^{-2}$
from the two energy denominators of the states containing pions,
$P^{-2}$ from four factors $1/\sqrt{2E_\pi}$, a factor $P^3$ from the
integral over the loop three-momentum, and finally $P^{-2}$ from the
two-nucleon energy denominator. Altogether, the graph (a) of Fig.\ 
\ref{fig2} gives a contribution of the order $P^0$ or equivalent
$Q^0$.

Thus we have to conclude that a complete calculation including all
terms of order $Q^\nu$ with $\nu\leq 0$ requires the inclusion of the
absorptive process. Indeed, already the imaginary contribution of the
absorption process, $\Im m\,E_d$ in (\ref{result}), turns out to be of
the same order of magnitude as the three-body contribution of order
$\nu =0$ calculated by Beane et al., $E^{tb,4}_d= -0.25\times
10^{-3}/m_{\pi^+}$ in the notation of \cite{beanenew}.

In summary, the contribution of the two-step process $\gamma
d\rightarrow pn\rightarrow\pi^0 d$ to the imaginary part of the
electric dipole (and magnetic quadrupole) amplitude for coherent pion
photoproduction on the deuteron has been calculated utilizing
unitarity constraints.  The result shows that this absorptive process
cannot be neglected in the extraction of the elementary neutron
amplitude.  It has been shown that a consistent $\chi$PT calculation
for $\gamma d\rightarrow\pi^0 d$ beyond next-to-leading order
requires indeed the inclusion of the absorptive process.

I thank H.\ Arenh\"ovel for useful discussions.  This work was
supported by the Deutsche Forschungsgemeinschaft (SFB 201).



\begin{references}

\bibitem{mainz} Mainz Microtron Proposal A1/1-96, Spokesperson: R.\ 
  Neuhausen.

\bibitem{koch} J.\ H.\ Koch and R.\ M.\ Woloshyn, Phys.\ Lett.\ B {\bf
    60}, 221 (1976); Phys.\ Rev.\ C {\bf 16}, 1968 (1977).

\bibitem{beane} S.\ R.\ Beane, C.\ Y.\ Lee, and U.\ van Kolck, Phys.\ 
  Rev.\ C {\bf 52}, 2914 (1995).

\bibitem{beanenew} S.\ R.\ Beane, V.\ Bernard, T.-S.\ H.\ Lee, Ulf-G.\ 
  Meissner, and U.\ van Kolck, Duke University Report No.\ 
  DUKE-TH-96-131, hep-ph/9702226, 1997.

\bibitem{ericson} T.\ Ericson and W.\ Weise, {\it Pions and Nuclei}
  (Clarendon Press, Oxford, 1988).

\bibitem{pilkuhn} H.\ M.\ Pilkuhn, {\it Relativistic Particle Physics}
  (Springer, New York, 1979).

\bibitem{hutcheon} D.\ A. Hutcheon {\it et al.,} Nucl.\ Phys.\ {\bf
    A535}, 618 (1991).

\bibitem{ha} For a review see, H.\ Arenh\"ovel and M.\ Sanzone, Few-Body
  Syst.\ Suppl.\ {\bf 3} (1991).

\bibitem{myletter} P.\ Wilhelm and H.\ Arenh\"ovel, Phys.\ Lett.\ B {\bf
    318}, 410 (1993).

\bibitem{meissner} V.\ Bernard, N.\ Kaiser, and Ulf-G.\ Meissner, Z.\ 
  Phys.\ C {\bf 70}, 483 (1996).

\bibitem{niskanen} See e.g., J.\ A.\ Niskanen, Nucl.\ Phys.\ {\bf
    A298}, 417 (1978); Phys.\ Lett.\ B {\bf 141}, 301 (1984).

\bibitem{wein1} S.\ Weinberg, Phys.\ Lett.\ B {\bf 251}, 288 (1990).

\bibitem{wein2} S.\ Weinberg, Nucl.\ Phys.\ {\bf B363}, 3 (1991).

\bibitem{wein3} S.\ Weinberg, Phys.\ Lett.\ B {\bf 295}, 114 (1992).

\bibitem{friar} T.\ D.\ Cohen, J.\ L.\ Friar, G.\ A.\ Miller, and U.\ 
  van Kolck, Phys.\ Rev.\ C {\bf 53}, 2661 (1996).
\end{references}
\end{document}